\documentclass[english,aps,prl,twocolumn,amsmath,amssymb,showpacs,superscriptaddress,notitlepage,longbibliography]{revtex4-2}
\usepackage{bm}
\usepackage[pdftex]{graphicx,hyperref}
\hypersetup{colorlinks = true, urlcolor = blue, linkcolor = blue, citecolor = blue}
\usepackage{mathtools}
\usepackage{ulem}
\usepackage{color}

\begin{document}

\title{Spin-Polarized Josephson Supercurrent in Nodeless Altermagnets}
%\title{Spin-Triplet Josephson Supercurrent in Nodeless Altermagnets}
%\title{Josephson Supercurrent Mediated by Pure Triplet Pairings in Nodeless Altermagnets}

\author{Chuang Li}
\thanks{These authors contributed equally.}
\affiliation{Center for Correlated Matter and School of Physics, Zhejiang University, Hangzhou 310058, China}
\affiliation{Hefei National Laboratory, Hefei, Anhui, 230088, China}

\author{Jin-Xing Hou}
\thanks{These authors contributed equally.}
\affiliation{Hefei National Laboratory, Hefei, Anhui, 230088, China}
\affiliation{International Center for Quantum Design of Functional Materials (ICQD), University of Science and Technology of China, Hefei, Anhui 230026, China}

\author{Fu-Chun Zhang}
\affiliation{Kavli Institute for Theoretical Sciences, University of Chinese Academy of Sciences, Beijing 100190, China}

\author{Song-Bo Zhang}
\email{songbozhang@ustc.edu.cn}
\affiliation{Hefei National Laboratory, Hefei, Anhui, 230088, China}
%\address{School of Emerging Technology, University of Science and Technology of China, Hefei, 230026, China}
\affiliation{International Center for Quantum Design of Functional Materials (ICQD), University of Science and Technology of China, Hefei, Anhui 230026, China}

\author{Lun-Hui Hu}
\email{lunhui@zju.edu.cn}
\affiliation{Center for Correlated Matter and School of Physics, Zhejiang University, Hangzhou 310058, China}

\begin{abstract}
Long-range propagation of equal-spin triplet Cooper pairs typically occurs in ferromagnet/$s$-wave superconductor junctions, where net magnetization plays a crucial role. Here, we propose a fundamentally different scenario in which Josephson supercurrents mediated exclusively by spin-triplet pairings emerge in systems with \textit{zero} net magnetization. We identify collinear altermagnets, particularly a subclass termed nodeless altermagnets, as ideal platforms to realize this phenomenon. These materials host spin-split Fermi surfaces that do not intersect altermagnetic nodal lines and support maximal spin-valley polarization, yielding fully spin-polarized electronic states at each valley. Consequently, Josephson junctions based on nodeless altermagnets sustain supercurrents solely through spin-polarized triplet pairing correlations, simultaneously contributed by spin-up Cooper pairs from one valley and spin-down Cooper pairs from the other. Furthermore, controlling the relative local inversion-symmetry breaking at the two interfaces enables a robust 0–$\pi$ transition without fine tuning, while adjusting the junction orientation allows a crossover between pure triplet and mixed singlet-triplet states. Our work thus establishes nodeless altermagnets as a unique platform for altermagnetic superconductors with magnetization-free spin-polarized supercurrents.
\end{abstract}

\maketitle

\textit{Introduction.--}
The recent discovery of collinear altermagnetism (AM) has significantly expanded our understanding of magnetic materials~\cite{vsmejkal2020crystal,Naka19NC,Ahn19PRB,hayami2019momentum,Yuan2020Giant,mazin2021prediction,ma2021multifunctional,ifmmode2022Beyond,ifmmode2022Emerging,krempasky2024altermagnetic,sdongFeSb2}. Unlike conventional antiferromagnets, AM hosts antiparallel spins coupled through crystalline symmetries such as rotation and reflection, establishing a new magnetic phase characterized by vanishing net magnetization and momentum-dependent spin splitting~\cite{jungwirth2024altermagnets,bai2024altermagnetism,fender2025altermagnetism,liu2025different}. This unconventional magnetic phase can be realized in diverse systems~\cite{maierPRB2023weak,he2023prl,bhowal2024prx,leeb2024prl,das2024prl,yuan2024prl,Roig2024prb,atasi2024prb,vsmejkal2024altermagnetic,duan2025prl,gu2025ferroelectric,zhu2025two}, and manifests a range of novel quantum phenomena including non-relativistic spin splitting~\cite{ifmmode2022Emerging}, crystal-symmetry-paired spin-valley locking (SVL)~\cite{ma2021multifunctional,Hu2025prx}, spin-orbital textures~\cite{wang2025prl,vila2024orbital}, and anomalous transport properties~\cite{Nakaprb,shao2021spin,FengZX22NE,Fernandestoplogical,zhang2024prl,RChen25PRL,Sudbo2025arXiv}. Recent experiments have observed both spin-splitting and SVL in various quantum materials~\cite{fedchenko2024ruo2,lin2024ruo2,gonzalez2023prl,krempasky2024MnTe,lee2024MnTe,osumi2024MnTe,liu2024chiral,reimers2024CrSb,ding2024CrSb,yang2025three,jiang2025metallic,zhang2025crystal}. While momentum-space spin splitting may also arise from mechanisms like spin-channel Pomeranchuk instabilities~\cite{Hirsch1990prb,wu2004prl,CJWu07PRB} or $d$-wave spin-density wave states~\cite{Ikeda1998prl}, SVL is unique to AMs thus far.

SVL represents a distinctive manifestation of spin-splitting under specific symmetry constraints~\cite{Hu2025prx}. When spin-orbit coupling is negligible, the spin-space group forbids spin-splitting along certain momentum directions. For example, the coexistence of symmetries $[{\cal C}_2||{\cal M}_{[11]}]$ and $[{\cal C}_2||{\cal M}_{[\bar{1}1]}]$ guarantee vanishing spin-splitting along the $k_x = \pm k_y$ directions, resulting in symmetry-protected altermagnetic nodal lines. Depending on whether these nodal lines intersect the Fermi surface, AMs can be divided into two classes: \textit{nodal} AMs and \textit{nodeless} AMs~\cite{jungwirth2024altermagnets}. This classification distinguishes different Fermi surface topologies and is relevant only for metallic phases. Particularly, nodeless AMs feature spin-split Fermi surfaces that avoid enclosing the $\Gamma$ point and inherently support SVL. Hence, SVL is a defining characteristic of nodeless AMs. While nodal AMs have been well explored~\cite{Ouassou2023prl,Zhang2024NatComm,Beenakker2023prb,Cheng2024prb,Banerjee2024prb,sun2023Andreev,Papaj2023Andreev,Wei2024prb,Sun2025prb,Chakraborty2025prl}, nodeless AMs present fundamentally distinct opportunities. In particular, SVL in AMs breaks time-reversal symmetry, a feature that remains underexplored but with great potential for spintronics applications and superconducting proximity effects.

\begin{figure}[t]
\centering
\includegraphics[width=\linewidth]{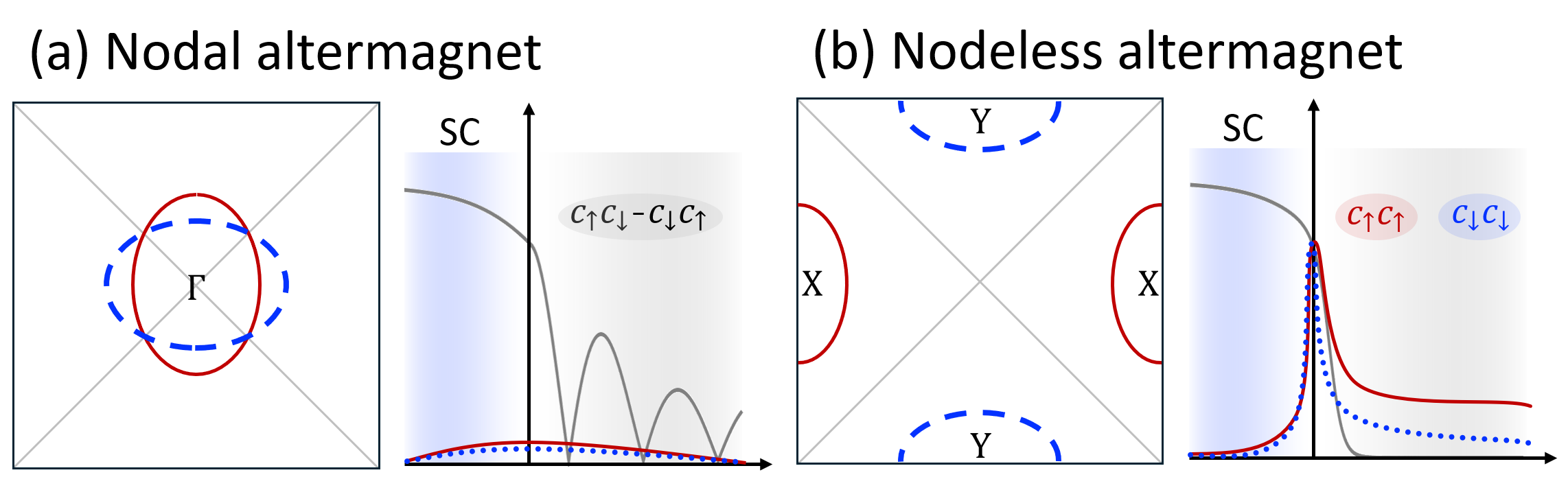}
\caption{Proximity effects in $s$-wave superconductor/AM junctions, showing dominant pairing correlations:
(a) Spin-singlet pairing with spatial oscillations in nodal AM metals~\cite{Ouassou2023prl,Zhang2024NatComm}.
(b) Spin-triplet pairing in nodeless AM metals, with $ c_\uparrow c_\uparrow $ and $c_\downarrow c_\downarrow$ contributed from two valleys, respectively.
Upper panels: Fermi surfaces with solid and dashed lines indicating spin-up and spin-down polarizations).
Lower panels: Corresponding pairing correlations across the junction.
}
\label{fig0}
\end{figure}

\begin{figure*}[t]
\centering
\includegraphics[width=0.99\linewidth]{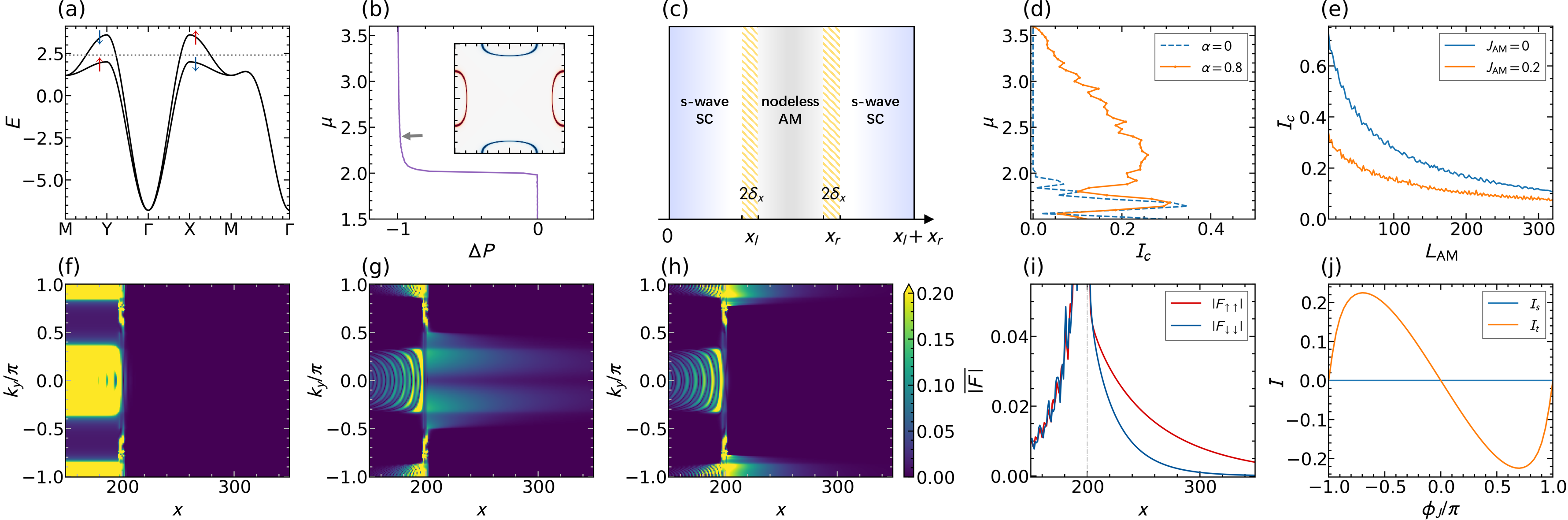}
\caption{Main results for $0^\circ$-aligned nodeless AM-based Josephson junctions.
(a) Band structure of the AM described by Eq.~\eqref{eq-ham0}, with red ($\color{red}\uparrow$) and blue ($\color{blue}\downarrow$) arrows denoting spin-polarized bands.
(b) Spin-valley polarization $\Delta P$ as a function of $\mu$ for bands in (a). Inset: spin-split Fermi surfaces at $\mu=2.4$ (dashed line in (a)).
(c) Illustration of the NSN Josephson junction geometry. The yellow region denotes the interface that breaks inversion symmetry.
(d) Critical Josephson current $I_c$ in the SC/AM/SC junction as a function of $\mu$, comparing spin-conserving ($\alpha=0$, blue dashed) and non-spin-conserving ($\alpha=0.8$, orange solid) interfaces.
(e) $I_c$ versus AM region length $L_{\text{AM}}$, with (orange) and without (blue) the AM spin-splitting term $J_{\text{AM}}$.
Proximity-induced pairing correlations near the SC/AM interface include:
(f) spin-singlet $|F_s(x,k_y)|$, 
(g) up-up triplet $|F_{\uparrow\uparrow}(x,k_y)|$, 
and (h) down-down triplet $|F_{\downarrow\downarrow}(x,k_y)|$.
(i) Spatial decay profiles of $k_y$-averaged equal-spin triplet pairing amplitudes: $|F_{\downarrow\downarrow}(x)|$ (red) and $|F_{\uparrow\uparrow}(x)|$ (blue), showing distinct decay rates.
(j) Current-phase relation $I(\phi_J)$ decomposed into singlet (blue) and triplet (orange) contributions, showing exclusively triplet-driven supercurrent.
Parameters: $t_1=1$, $t_2=0.7$, $\mu=2.4$, $J_\text{AM}=0.2$, $\Delta_0=0.02$, $\alpha=0.8$, $T=0.02$, and a small frequency $\omega=0.02$. $(L_\text{SC},L_\text{AM})=(200,1600)$ in the SC/AM junction, $(L_\text{SC},L_\text{AM},L_\text{SC})=(50,50,50)$ in the SC/AM/SC junction.
}
\label{fig:valley}
\end{figure*}

In this work, we demonstrate nodeless AM as ideal platforms for proximity-induced \textit{pure} spin-triplet correlations without invoking net magnetization. For AM-based Josephson junctions, Fig.~\ref{fig0}(a) shows that nodal AMs permit spatially oscillating spin-singlet pairing~\cite{Ouassou2023prl,Zhang2024NatComm}. In contrast, nodeless AMs uniquely generate spin-polarized triplet correlations containing both $c_{\uparrow}c_{\uparrow}$ (from $X$-valley) and $c_{\downarrow}c_{\downarrow}$ (from $Y$-valley) [Fig.~\ref{fig0}(b)]. This valley degree of freedom facilitates an experimentally tunable, orientation-dependent Josephson effect, allowing controlled crossover from pure triplet to mixed singlet-triplet states. Tuning the relative local inversion-symmetry breaking at the two junction interfaces triggers a robust 0–$\pi$ transition without fine tuning. Our work provides an extrinsic mechanism yielding exotic altermagnetic superconductors with spin-polarized supercurrents that break spin-space group symmetries~\cite{parshukov2025exotic}.

\textit{Spin-valley polarization in AMs.--}
To quantify SVL in AMs, we introduce the spin-valley polarization $\Delta P$. Consider a system with two inequivalent valleys, $X$ and $Y$. The spin polarization around each valley is $P_{X(Y)} = ({\cal N}_{X(Y),\uparrow} - {\cal N}_{X(Y),\downarrow})/({\cal N}_{X(Y),\uparrow} + {\cal N}_{X(Y),\downarrow})$, where ${\cal N}_{X(Y),\sigma}$ denotes the spin-resolved density of states at valley $X$ or $Y$ for spin $\sigma \in \{\uparrow,\downarrow\}$. The spin-valley polarization is then defined by $\Delta P \equiv P_X P_Y$. In normal metals with spin-degenerate bands, the spin polarization vanishes, resulting in $\Delta P=0$. In contrast, AMs feature vanishing net magnetization, which holds $P_X + P_Y=0$, leading to $\Delta P < 0$. While $\Delta P$ can take any value in between $-1$ and $0$, it becomes quantized to $-1$ when each valley hosts fully spin-polarized Fermi surfaces. This represents perfect SVL, achievable in materials such as Rb$_{1-\delta}$V$_2$Te$_2$O and KV$_2$Se$_2$O~\cite{jiang2025metallic,zhang2025crystal}. In this work, we focus on this maximal $\Delta P$ scenario and demonstrate that it generates purely spin-polarized triplet pairing in the bulk AM via proximity effect. As a proof of concept, we study a minimal $d$-wave AM Hamiltonian on a square lattice:
\begin{align} \label{eq-ham0}
{\cal H}_{\text{AM}}(\bm{k}) = \epsilon_0(\bm{k}) \hat{\sigma}_0 - 2J_{\text{AM}}(\cos k_x - \cos k_y) \hat{\sigma}_z ,    
\end{align}
where $\epsilon_0(\bm{k}) = -2t_1(\cos k_x + \cos k_y) - 4t_2\cos k_x \cos k_y$, $\hat{\sigma}_{0}$ and $\hat{\sigma}_{\nu}$ (with $\nu=x,y,z$)  are the identity and Pauli matrices acting on spin space, $t_1$ ($t_2$) is the nearest (next-nearest) neighbor hopping amplitude, and $J_{\text{AM}}$ denotes the strength of the $d$-wave altermagnetic spin-splitting. The resulting band structure exhibits valley-dependent splittings [Fig.~\ref{fig:valley}(a)]: a positive splitting $+4J_{\text{AM}}$ at $X$ while a negative splitting $-4J_{\text{AM}}$ at $Y$. Thus, for a wide range of chemical potentials (around $\mu \sim 2.4$), the system hosts fully spin-polarized Fermi pocket centered at each valley [Fig.~\ref{fig:valley}(b)], achieving maximal spin-valley polarization ($\Delta P = -1$). Note that this scenario is not limited to this specific model but is achievable on various lattice systems~\cite{chakraborty2024strain,zhu2025design,li2025pressure}.

\textit{Josephson junctions based on nodeless AMs.--}
We next explore the role of maximal spin-valley polarization in SC/AM/SC Josephson junctions [Fig.~\ref{fig:valley}(c)]. We consider a planar junction formed by two $s$-wave superconductors (SCs) separated by a nodeless AM. For junctions oriented along the $x$-direction, the two valleys in the AM become fully decoupled, and each behaves like a half-metal. This effectively creates two parallel half-metallic transport channels that naturally carry spin-polarized triplet Josephson currents. To analyze the Josephson effect, we model the junction with the Hamiltonian
\begin{align}\label{eq-model-ham}
{\cal H}_{\text{SNS}} = \sum_{k_y}({\cal H}_0 + {\cal H}_\text{L1} + {\cal H}_\text{AM} + {\cal H}_\text{L2} + {\cal H}_\text{SC-AM}),
\end{align}
in Nambu basis $C_x^\dagger = (c_{x\uparrow}^\dagger, c_{x\downarrow}^\dagger, c_{x\uparrow}, c_{x\downarrow})$. We assume translation symmetry along the interface, so $k_y$ remains a good quantum number. The kinetic term, ${\cal H}_0= \sum_{x} \{ C_x^\dagger ( -2t_1 \cos{k_y} -\mu ) \hat{\tau}_z\hat{\sigma}_0 C_x +[C_{x+1}^\dagger ( -t_1 -2t_2 \cos{k_y} ) \hat{\tau}_z\hat{\sigma}_0 C_x +\text{h.c.}]\}$, acts throughout the entire system. $\hat{\tau}_{\nu}$ ($\nu\in\{x,y,z\}$) are the Pauli matrices in Nambu space. The on-site $s$-wave pairing terms in the two superconducting leads are ${\cal H}_\text{L1} = -\Delta_0 \sum_{0\leq x < x_l} C_x^\dagger \hat{\tau}_y\hat{\sigma}_y C_x$ with $x_l=L_{\text{SC}}$ and ${\cal H}_\text{L2}= -\Delta_0\sum_{x_r \leq x < x_r+L_{\text{SC}}} C_x^\dagger [\cos\phi_J\hat{\tau}_y+\sin\phi_J\hat{\tau}_x] \hat{\sigma}_y C_x$ with $x_r=x_l+L_{\text{AM}}$, where $\Delta_0$ is the pairing gap and $\phi_J$ is the superconducting phase difference. The term for nodeless AM in the junction reads ${\cal H}_\text{AM}= J_\text{AM} \sum_{x_l\leq x< x_r}[C_{x}^\dagger (2\cos{k_y}) \hat{\tau}_z\hat{\sigma}_z C_x - (C_{x+1}^\dagger \hat{\tau}_z\hat{\sigma}_z C_x +\text{h.c.})]$. The interfacial Rashba spin-orbit coupling, arising from structural inversion symmetry breaking~\cite{asano2021andreev}, is confined to a finite interfacial region, i.e., ${\cal H}_\text{SC-AM}=\alpha \sum_{|x-x_{l/r}| \leq \delta_x} \{ C_{x}^\dagger (-\sin{k_y})\hat{\tau}_0\hat{\sigma}_x C_x + [C_{x+1}^\dagger \tfrac{i}{2}\hat{\tau}_z\hat{\sigma}_y C_x + \text{h.c.}] \}$. While we use $\delta_x=2$ below, our main conclusions remain unaffected by this choice.

Based on the continuity equation [see Sec.~S1 in Supplementary Material (SM)~\cite{sm2025}], we calculate the local supercurrent flowing across the junction as~\cite{Asano2001PRB,Sakurai17PRB,SBZhang20PRB},
\begin{align} \label{eq-junction-Ic}
{I}_x(\phi_J) =& -\frac{4e}{\hbar\beta} \sum_{\omega,k_y} \text{Im} \left[ \text{Tr} [ \hat{T}_h^\dagger F_{x+1} \hat{T}_e \tilde{F}_{x} ] \right],
\end{align}
where $\beta = 1/k_B T$, $\omega=(2n+1)\pi/\beta$ are Matsubara frequencies, $\hat{T}_{e/h}(k_y)$ are electron (hole) hopping matrices, $F_{x}(\omega,k_y)$ is the anomalous Green's function at site $x$ inside the AM, and $\tilde{F}_{x}(\omega,k_y)$ is the surface anomalous Green's function~\footnote{The surface Green’s function $\tilde{F}_{x}(\omega,k_y)$ at $x$ site is defined as $1/(i\omega_n - {\cal H}_\text{surf})$, where ${\cal H}_\text{surf}$ describes the subsystem Hamiltonian for all sites satisfying $x' \leq x$.}. The current is uniformity within the AM, i.e.,~$I_x = I$ throughout the junction. The critical current $I_c$ is defined as the extreme value of $I(\phi_J)$ within $-\pi < \phi_J < 0$.

Figure~\ref{fig:valley}(d) presents $I_c$ as a function of chemical potential $\mu$. We set $\Delta_0=0.02t_1$, corresponding to a coherence length  $\xi_{\text{SC}}\approx 160$. For $\alpha = 0$, $I_c$ vanishes for $L_{\text{AM}} \geq 16$ ($\sim 0.1 \xi_{\text{SC}}$) for $\mu$ in the maximal spin-valley polarization region. This suppression stems from spin $U(1)$ symmetry, which restricts the junction to spin-singlet pairing correlations that decay rapidly in the nodeless AM. However, introducing a finite $\alpha$-term breaks spin-rotation symmetry at the interfaces, enabling singlet-to-triplet conversion and leading to finite $I_c$ [solid orange line for $\alpha=0.8$, Fig.~\ref{fig:valley}(d)]. Alternatively, such conversion can be achieved using spin-orbit-coupled $s$-wave SCs or interfacial spin-canting [see Sec.~S2 in SM~\cite{sm2025}]. Remarkably, at $\alpha = 0.8$, a significant $I_c$ emerges for both short and long junctions (e.g.,~$L_{\text{AM}} = 300 \sim 2\xi_{\text{SC}}$), with magnitude comparable to the nonmagnetic counterpart ($J_{\text{AM}}=0$) [Fig.~\ref{fig:valley}(e)].

\textit{Proximity-induced pure triplet correlations.--}
Equation~\eqref{eq-junction-Ic} demonstrates that on-site pairing correlations fully govern the supercurrent. To understand the microscopic origin of the pronounced $I_c$, we analyze the proximity-induced pairing correlations within the nodeless AM, extracted from $F_{x}(\omega,k_y)$ in Eq.~\eqref{eq-junction-Ic}. These correlations decompose as
\begin{align} \label{eq-pair-correlation}
F =& -i\hat{\sigma}_y F_s +\tfrac{\hat{\sigma}_0+\hat{\sigma}_z}{2} F_{\uparrow\uparrow} +\tfrac{\hat{\sigma}_0-\hat{\sigma}_z}{2} F_{\downarrow\downarrow} +\hat{\sigma}_x F_z,
\end{align}
where $F_{s/z}= (F_{\downarrow\uparrow} \mp F_{\uparrow\downarrow})/2$. $F_s$ corresponds to the singlet pairing, while $F_z$, $F_{\uparrow\uparrow}$, and $F_{\downarrow\downarrow}$ represent the triplet pairings. To clarify the behavior of induced pairings, it is constructive to first examine the simpler NS junction setup. As shown in Fig.~\ref{fig:valley}, only spin-triplet correlations exhibit long-range proximity effects in the AM. Specifically, the singlet component $|F_s|$ decays rapidly for all $k_y$ [Fig.~\ref{fig:valley}(f)]. In sharp contrast, the equal-spin triplet components $|F_{\uparrow\uparrow}|$ and $|F_{\downarrow\downarrow}|$, which arise at finite frequencies, penetrate deeply into the AM [Figs.~\ref{fig:valley}(g-h)]. These proximity-induced triplet pairings exhibit the same symmetry-breaking characteristics as the bulk AM and thus can be classified as extrinsic altermagnetic SCs~\cite{parshukov2025exotic}. As a result, they are fundamentally distinct from the $p$-wave triplet states that arise from altermagnetic fluctuations~\cite{wu2025intra}.

The proximity-induced triplet pairing is intrinsically spin-polarized, manifested in distinct decay rates of the $k_y$-averaged $|F_{\uparrow\uparrow}|$ and $|F_{\downarrow\downarrow}|$ [Fig.~\ref{fig:valley}(i)]. This polarization originates from valley-locked pairing correlations: $|F_{\uparrow\uparrow}|$ emerges exclusively from the $X$-valley Fermi surface with $v_{F,\uparrow} \approx 3.3$, while $|F_{\downarrow\downarrow}|$ stems solely from the $Y$-valley with $v_{F,\downarrow} \approx 1.2$. These velocities qualitatively determine the decay lengths $\lambda_\sigma$ (via $\lambda_\sigma \propto v_{F,\sigma}$) that yield fits to the decay profiles using $|F_{\sigma\sigma}| \propto \tfrac{1}{x} e^{-x/\lambda_\sigma}$ for clean systems, directly governing the observed decay asymmetry in Fig.~\ref{fig:valley}(i). We demonstrate that the valley-spin locked pairing directly encodes maximal spin-valley polarization ($\Delta P = -1$), and confirm vanishing finite-size magnetization for large $L_{\text{AM}} \gg 1/k_F$~\cite{LHHu25SCPMA,Hodt2024prb}. Thus, these pairing correlations naturally drive the magnetization-free spin-polarized Josephson supercurrent. Since the hopping matrices $\hat{T}_{e/h}$ in Eq.~\eqref{eq-junction-Ic} are diagonal in spin space, the supercurrent $I(\phi_J)$ decomposes as,
\begin{align} \label{eq-Ij-st}
I(\phi_J) = I_{s}(\phi_J) + I_{t}(\phi_J) + I_{st}(\phi_J),
\end{align}
where $I_s\propto F_s\tilde{F}_s$ and $I_t\propto F_{\uparrow\uparrow}\tilde{F}_{\uparrow\uparrow} + F_{\downarrow\downarrow}\tilde{F}_{\downarrow\downarrow} + F_z\tilde{F}_z$ [see Sec.~S1 of SM~\cite{sm2025}]. Due to the pure triplet pairings, the singlet-triplet mixing contribution ($I_{st}$) to $I$ vanishes. As shown in Fig.~\ref{fig:valley}(j), $I_s$ vanishes for all $\phi_J$, leaving triplet correlations as the sole source of supercurrent. The triplet supercurrent polarization ratio is $I_{t,\uparrow\uparrow} / I_{t,\downarrow\downarrow} \approx 3.4$ for our parameters [see Sec.~S3 in SM~\cite{sm2025}]. This polarization can be enhanced by tuning the ratio $v_{F,\uparrow}/v_{F,\downarrow}$. Our results indicate the spin-polarized supercurrent feature of altermagnetic SCs.

\begin{figure}[t]
\centering
\includegraphics[width=0.95\linewidth]{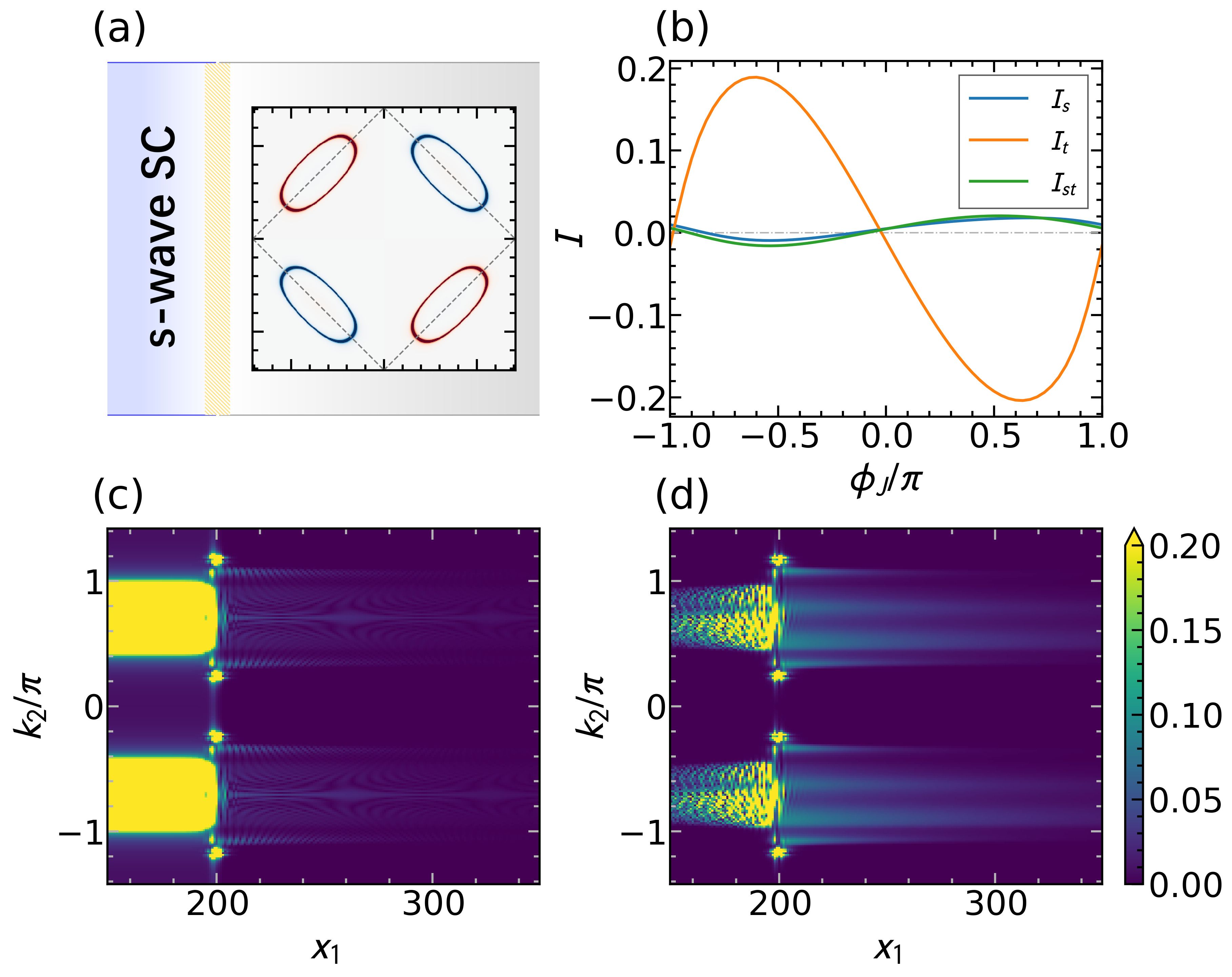}
\caption{(a) Sketch of the SC/AM junction. Inset: Fermi surface of the AM rotated by $\pi/4$, with gray dashed line marking the first Brillouin zone boundary.
(b) Phase-dependent Josephson currents: singlet ($I_s$), triplet ($I_t$), and mixed ($I_{st}$) contributions versus phase difference $\phi_J$ in the SC/AM/SC junction.
Spatial evolution of $|F_s(k_y)|$ in (c) and $|F_{\uparrow\uparrow}(k_y)|$ in (d) near the SC/AM interface.
}
\label{fig:rot45}
\end{figure}

\textit{Effects of junction orientation.--}
The robustness of these results originates from the forbidden inter-valley spin-singlet pairing channel. We now show how junction orientation controls the emergence of this channel. In a $45^\circ$-aligned junction [Fig.~\ref{fig:rot45}(a), see Sec.~S2 in SM~\cite{sm2025}], the global Fermi surface rotation hybridizes the $X$ and $Y$ valley indices in the rotated $k_1$-$k_2$ frame. While $k_2$ remains conserved, this mixing enables both intra-valley (e.g.,~$\langle c_{X,\uparrow}c_{X,\uparrow} \rangle$) and inter-valley (e.g.,~$\langle c_{X,\uparrow}c_{Y,\downarrow}\rangle$) pairings via proximity. In Fig.~\ref{fig:rot45}(b), we find both $I_s(\phi_J)$ and $I_t(\phi_J)$ contribute to the supercurrent, while the spin-singlet correlation persists even at $\alpha=0$ [see Sec.~S4 in SM~\cite{sm2025}]. For $\alpha \neq 0$, Figs.~\ref{fig:rot45}(c-d) show coexisting singlet and triplet correlations throughout the nodeless AM. Thus, rotating the junction orientation from $0^\circ$ to $45^\circ$ induces a crossover from pure triplet to mixed singlet-triplet supercurrent, although triplet pairing remains dominant. This orientation dependence directly manifests the anisotropic spin-splitting inherent to AM, fundamentally distinguishing AM-based junctions from other systems such as half metals~\cite{Buzdin2005rmp}.

\textit{Tunable $0$-$\pi$ transition.--}
The Josephson current mediated by triplet pairings can be controlled to realize tunable $0$-$\pi$ transition. To illustrate this, we first analyze how the interfacial spin-orbit coupling affects the pairing correlations in the NS junction. To incorporate valley degrees of freedom, we compute the $k_y$-summed pairing correlations, $F_i(\omega) = 1/2\pi \int_{-\pi}^{\pi} dk_y F_i(\omega,k_y)$, where $F_i$ (with $i \in \{s, z, \uparrow\uparrow, \downarrow\downarrow \}$) are defined in Eq.~\eqref{eq-pair-correlation}.

\begin{figure}[t]
\centering
\includegraphics[width=0.98\linewidth]{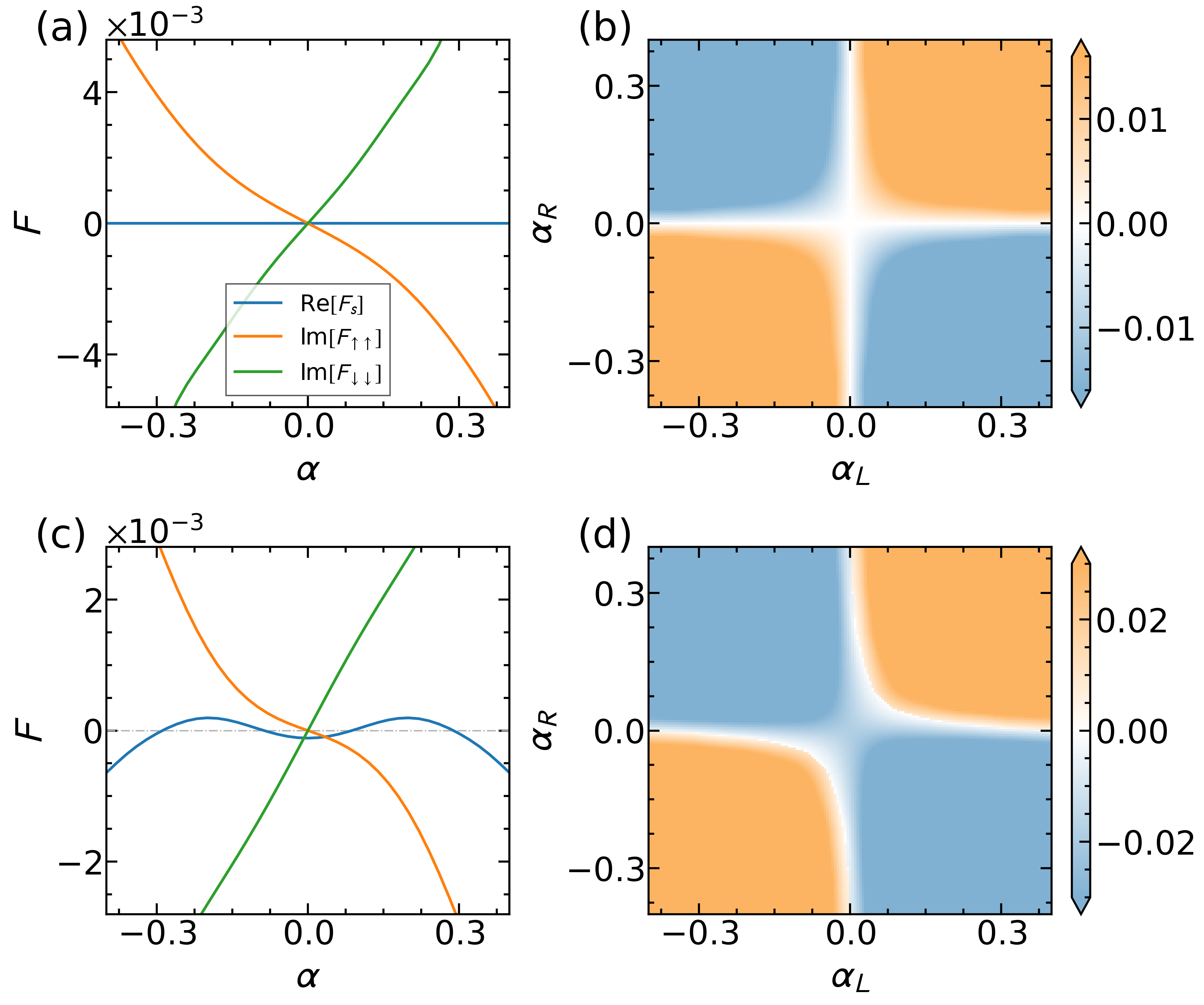}
\caption{(a) Dependence of pairing correlations on interfacial Rashba strength $\alpha$ in the $0^\circ$-junction.
(b) Critical current phase diagram: $I_c$ versus interfacial Rashba couplings $\alpha_L$ and $\alpha_R$ in the SC/AM/SC junction.
(c) Dependence of pairing correlations on interfacial Rashba strength $\alpha$ in the $45^\circ$-junction.
(d) Critical current $I_c(\alpha_L,\alpha_R)$ exhibiting a butterfly pattern in the $45^\circ$-junction.
}
\label{fig:current}
\end{figure}

For the $0^\circ$-junction, Fig.~\ref{fig:current}(a) shows $F_s$, $F_{\uparrow\uparrow}$, and $F_{\downarrow\downarrow}$ as functions of $\alpha$ at $x = L_{\text{SC}} + 20$, deep within the AM bulk. While $F_s$ vanishes at this distance ($20 > 0.1\xi_{\text{SC}}$), both triplet components $F_{\uparrow\uparrow}$ and $F_{\downarrow\downarrow}$ emerge and grow with $|\alpha|$. Notably, the triplet components reverse sign when $\alpha$ changes sign, $\alpha \to -\alpha$, indicating their sensitivity to the sign of the interfacial spin-orbit coupling. Extending to the SNS junction, we introduce two independent interfacial spin-orbit coupling strengths $\alpha_L$ and $\alpha_R$ at the left and right interfaces, respectively. Remarkably, by tuning these two parameters, we observe robust $0$-$\pi$ transitions in the Josephson current [Fig.~\ref{fig:current}(b)]. The phase boundaries lie along the $\alpha_L=0$ and $\alpha_R=0$, intersecting to form a cross (+) symbol. Crucially, it follows $\text{sign}[I_c] = \text{sign}[\alpha_L \alpha_R]$, demonstrating that the relative sign of spin-orbit coupling at the interfaces determines whether the junction is in the $0$- or $\pi$-state.

In the $45^\circ$-junction, $F_s$ coexists with $F_{\uparrow\uparrow}$ and $F_{\downarrow\downarrow}$ [Fig.~\ref{fig:current}(c)]. Since $F_s$ is an even function of $\alpha$, however, the sharp 0-$\pi$ transition lines at $\alpha_L=\alpha_R=0$ become avoided crossings. Consequently, this transforms the $0$-$\pi$ boundaries in the $\alpha_L$-$\alpha_R$ plane into a distinctive butterfly pattern [Fig.~\ref{fig:current}(d)].

\textit{Discussions and conclusions.--}
Finally, we note that the spin-triplet Josephson current $I_t$ contains contributions from both even-$\omega$ and odd-$\omega$ triplet correlations, as shown in the $\omega$-summation in Eq.~\eqref{eq-junction-Ic}. Explicitly, the triplet pairing can be decomposed as $F_{\uparrow\uparrow(\downarrow\downarrow)}= F_{\uparrow\uparrow(\downarrow\downarrow)}^{\text{even}} + F_{\uparrow\uparrow(\downarrow\downarrow)}^{\text{odd}}$ with $F_{\uparrow\uparrow(\downarrow\downarrow)}^{\text{even}}(\omega) = F_{\uparrow\uparrow(\downarrow\downarrow)}^{\text{even}}(-\omega)$ and $F_{\uparrow\uparrow(\downarrow\downarrow)}^{\text{odd}}(\omega)=-F_{\uparrow\uparrow(\downarrow\downarrow)}^{\text{odd}}(-\omega)$~\cite{Chakraborty2022prl}. Accordingly, $I_t$ in Eq.~\eqref{eq-Ij-st} can be separated into two parts,
\begin{align}
I_t(\phi_J) = I_t^{\text{o}}(\phi_J) + I_t^{\text{e}}(\phi_J),
\end{align}
with $I_t^{\text{e(o)}} \propto F_{\uparrow\uparrow}^{\text{even(odd)}} \tilde{F}_{\uparrow\uparrow}^{\text{even(odd)}} + F_{\downarrow\downarrow}^{\text{even(odd)}} \tilde{F}_{\downarrow\downarrow}^{\text{even(odd)}}$. Hence, $I_t^{\text{o}}$ can serve as a direct detector for odd-$\omega$ triplet pairing when $I_s(\phi_J)=I_t^{\text{e}}(\phi_J)=0$. As shown in Sec.~S3 of SM~\cite{sm2025}, the interfacial spin-orbit coupling induces coexisting even-$\omega$ and odd-$\omega$ triplets, yielding finite $I_t^{\text{o}}$ and $I_t^{\text{e}}$. In contrast, an interfacial spin-canting, described by $\mathcal{H}_\text{SC-AM} \propto \sum_{|x-x_{l/r}| \leq \delta_x} C_x^\dagger [\cos (x\pi/2) \hat{\tau}_z\hat{\sigma}_{x}+\sin (x\pi/2) \hat{\tau}_0\hat{\sigma}_{y}] C_x$, produces purely odd-$\omega$ triplets, resulting in only $I_t^{\text{o}}$. This provides a direct signature of odd-$\omega$ spin-triplet pairing in Josephson current measurements.

In summary, we have shown that nodeless altermagnets with maximal spin-valley polarization provide a unique platform for generating pure spin-triplet Josephson currents without net magnetization. The valley-locked pairing mechanism, in which two equal-spin triplet pairing correlations originate exclusively from two separated valleys, respectively, enables long-range triplet proximity effects unattainable in conventional metals or nodal altermagnets. Crucially, this system exhibits two experimentally tunable control knobs: (i) junction orientation, which governs the triplet purity and enables a crossover from exclusive triplet supercurrents ($0^\circ$-junction) to hybrid singlet-triplet states ($45^\circ$-junction); and (ii) interfacial symmetry breaking ($\alpha_L$ and $\alpha_R$), which triggers robust $0$-$\pi$ transitions without fine tuning, following the sign rule $\text{sign}[I_c] = \text{sign}[\alpha_L \alpha_R]$.

Experimental realization of our proposal is feasible using well-established fabrication techniques with spin-valley-locked altermagnets, such as KV$_2$Se$_2$O~\cite{jiang2025metallic}, Rb$_{1-\delta}$V$_2$Te$_2$O~\cite{zhang2025crystal}, and SrFe$_4$O$_{11}$~\cite{wan2024high}. Notably, the predicted spin-triplet Josephson supercurrent exhibits exceptional robustness against Zeeman fields [see Sec.~S5 in SM~\cite{sm2025}], which provides a distinctive signature contrasting sharply with singlet-dominant supercurrent. Our findings thereby establish nodeless altermagnets as a functional material platform for \textit{magnetization-free superconducting spintronics}. Combining our results with prior studies~\cite{Ouassou2023prl,Zhang2024NatComm,Beenakker2023prb,Cheng2024prb,Banerjee2024prb,sun2023Andreev,Papaj2023Andreev,Wei2024prb,Sun2025prb,Chakraborty2025prl} yields a comprehensive framework for superconducting proximity effects in altermagnets, thereby establishing the theoretical basis for exotic altermagnetic superconductors~\cite{parshukov2025exotic}.

\begin{acknowledgments}
We thank H.~K.~Jin and C.~X.~Liu for helpful discussions. 
C.L. and L.H.H. were supported by the start-up fund of Zhejiang University and the Fundamental Research Funds for the Central Universities (Grant No. 226-2024-00068). 
C.L. was also supported by central fiscal special-purpose fund (Grant No.2021ZD0302500).
J.X.H. and S.B.Z. were supported by the start-up fund at HFNL, and the Innovation Program for Quantum Science and Technology (Grant No. 2021ZD0302801).
\end{acknowledgments}

\bibliography{Refsdata}
\end{document}